\documentclass[11pt]{article}
\usepackage{hyperref}
\pdfoutput=1
\begin{document}
\title{Siquieros accidental painting technique: a fluid mechanics point of view}
\author{Sandra Zetina$^1$ and Roberto Zenit$^2$ \\
$^{1.}$ Instituto de Investigaciones Est\'eticas \\  $^{2.}$Instituto de Investigaciones en Materiales \\ Universidad Nacional Aut\'onoma de M\'exico \\
Cd. Universitaria, M\'exico D.F., 04510 \\ M\'EXICO} \maketitle
\begin{abstract}
This is an entry for the Gallery of Fluid Motion of the 65st
Annual Meeting of the APS-DFD ( fluid dynamics video ). This
video shows an analysis of the `accidental painting' technique developed by D.A. Siqueiros, a famous Mexican muralist. We reproduced the technique that he used: pouring layers of paint of different colors on top of each other. We found that the layers mix, creating aesthetically pleasing patterns, as a result of a Rayleigh-Taylor instability. Due to the pigments used to give paints their color, they can have different densities. When poured on top of each other, if the top layer is denser than the lower one, the viscous gravity current undergoes unstable as it spread radially. We photograph the process and produced slowed-down video to visualize the process.

\end{abstract}

\section{Introduction}

In the spring of 1936, the famous Mexican muralist David Alfaro Siqueiros \cite{Siqueiros} organized an experimental painting workshop in New York: a group of artists focused in developing painting techniques through empirical experimentation of modern and industrial materials and tools. Among the young artists attending the workshop was Jackson Pollock \cite{Pollock}. They tested different lacquers and a number of experimental techniques. One of the techniques, named by Siqueiros as a ``controlled accident,'' consisted in pouring layers of paint of different colors on top of each other. After a brief time, the paint from the lower layer emerged from bottom to top creating a relatively regular pattern of blobs. This technique led to the creation of explosion-inspired textures and catastrophic images. We conducted an analysis of this process. We experimentally reproduced the patterns ``discovered'' by Siqueiros and analyzed the behavior of the flow. We found that the flow is driven by the well-known Rayleigh Taylor instability \cite{Instabilities}: different colors paints have different densities; a heavy layer on top of a light one is an unstable configuration. The blobs and plumes that result from the instability create the aesthetically pleasing patterns. We discuss the importance of fluid mechanics in artistic creation.

\section{Experimental Conditions}
We used the same type of paints that Siqueiros used. Their physical properties are shown in Table \ref{Table_props}. With these combinations, both the Reynolds and Artwood numbers are small. The paint layers where dripping on top of a horizontal glass sheet. Volumes of approximately 50 and 25 ml were deposited for the bottom and bottom layers, respectively. The flow was photographed with a computer-controlled HD digital camera (FinePix Si Pro, Fujifilm), such that long term time sequences could be obtained. The time interval between photos was either 500 or 143 ms.

\begin{table}

\begin{tabular}{|c|c|c|c|c|}
  \hline
 Fluid  & density, kg/m$^3$ & viscosity Pa & Ar & Re \\
  \hline
  white paint & 1110 & 2.5 & 5.1$\times10 ^{-2}$ & 6.9$\times10 ^{-5}$ \\
  black paint & 1002 & 11.7 &  &  \\
  \hline
  yellow paint & 1080 & 3.6 & 3.4$\times10 ^{-2}$ & 1.1$\times10 ^{-4}$ \\
  transparent lacquer & 1008 & 12.9 &  &  \\
  \hline
\end{tabular}
  \centering
  \caption{Properties of the fluids and fluid combinations used. $Ar=(\rho_1-\rho_2)/(\rho_1+\rho_2)$, is the Artwood number and $Re=U_f H \rho/\mu$, is the Reynolds number, where $U_f$ and $H$ and the velocity of the front and the thickness of the layer, respectively.}\label{Table_props}
\end{table}

To our knowledge an formal analysis of this process has not been studied to date.

\section{Videos}

Our video contributions can be found at:

\begin{itemize}
    \item \href{http://somewhere.net}{Video 1, mpeg4, full
resolution}

    \item \href{http://somewhere.net}{Video 2, mpeg2, low
resolution}

\end{itemize}


\begin{thebibliography}{99}

\bibitem{Siqueiros}
P. Stein, \emph{Siqueiros: His Life and Works}. (Intl. Pub., 1994).

\bibitem{Pollock}
E. G. Landau, \emph{Jackson Pollock}. (Abrams, 2010).

\bibitem{Instabilities}
F. Charru, \emph{Hydrodynamic Instabilities}. (Cambridge University Press, 2011).

\end{thebibliography}
\end{document}